\documentclass[preprint,floats,superscriptaddress,usenames]{revtex4}

\usepackage{amsmath}
\usepackage{graphicx}
\newcommand{\trr}[1]{{#1}}

\newcommand{\rr}{{\bf r}}

\begin{document}
\date{\today}

\title{Steric modulation of ionic currents in DNA translocation through nanopores}

\author{Valerio Mazzone}

\address{Dipartimento di Fisica, Universit\`a La Sapienza, P.le A. Moro 2, 00185, 
Rome, Italy}

\author{Simone Melchionna}

\address{Istituto Processi Chimico-Fisici, Consiglio Nazionale delle Ricerche, Italy}

\author{Umberto Marini Bettolo Marconi}

\address{Scuola di Scienze e Tecnologie, 
Universit\`a di Camerino, Via Madonna delle Carceri, 62032, Camerino, INFN Perugia, Italy}


\begin{abstract}
Ionic currents accompanying DNA translocation strongly depend on 
molarity of the electrolyte solution and 
the shape and surface charge of the nanopore. 
By means of the Poisson-Nernst-Planck equations it is shown 
how conductance is modulated by the presence of the DNA intruder 
and as a result of competing electrostatic and confinement factors.
The theoretical results reproduce quantitatively the experimental ones 
and are summarized in a conductance diagram that allows 
distinguishing the region of reduced conductivity from the region of enhanced conductivity 
as a function of molarity and the pore dimension.
\end{abstract}

\maketitle

\section{Introduction}

Translocation of DNA in narrow pores plays a central role in many biological 
processes, such as viral infection by phages and interbacterial DNA transduction
\cite{molbio} as much as in the development of novel devices for high-throughput and low-cost
biotechnological applications \cite{dekkergen,venkatesan2011nanopore}.
In the last decade, several experimental studies have explored the translocation 
process of DNA through protein channels across cellular membranes or microfabricated channels  \cite{vlassiouk2008ionic}.

Solid state nanopores have attracted considerable attention as ideal devices for
reading off the base pairs while tracking the DNA motion through 
nanopores \cite{brantongen}.  
 Experimentally, during the DNA translocation through a nanopore one measures rapid variations
 of the ionic currents due both to the steric hindrance and the negative DNA charge density distribution.
 In principle, these variations can be used to decypher the DNA base pair sequence, if one is able to 
 finely identify the factors which modulate the current.  
To this purpose, one must  characterize the differences between the ionic signal when the DNA is in 
the pore or away from it.
One observes that the ionic current shows a larger/lower value when the DNA is inside/outside the pore
depending on the concentration of the electrolyte.
Since the reservoirs provide an amount of counterions sufficient to screen the DNA charge
the number of charge carriers increases and thus determines a larger conductance, 
in spite of the fact that the effective section available to the passage of ions  is smaller.

From the theoretical point of view
 DNA translocation and the accompanying electrokinetic transport of electrolytes
involve the comprehension of the competition between electrostatic, excluded volume and fluid-atom hydrodynamic 
couplings, with the key role played by the local confinement. 
Therefore, understanding the physical mechanisms that regulate ionic transport calls for an 
accurate determination of each contribution and, wherever possible, for realistic 
computational modeling.
In addition, as the diameters of the pore and of DNA can be as small as a few nanometers, 
ionic transport is genuinely microscopic. At such length scale a careful evaluation of the
local interactions requires a sophisticated theoretical treatment \cite{marconi2013effective}.

When DNA translocates in a solid state pore, it assumes an elongated conformation, 
due to both the electrostatic repulsion stemming from different parts of the DNA backbone, 
and by the axisymmetric and narrow shape of the pore.
One of the most interesting effects of the high confinement is that the
ensuing ionic current can be modulated in different and opposing ways. In fact, 
several authors have reported that ionic current can be either enhanced or blocked 
by the presence of DNA within the pore \cite{meller2001voltage,fologea2005detecting,mara2004asymmetric,storm2005fast,cao2012translocation,gupta2014dna}.
Current blockage refers to the fact that ionic current could be temporarily reduced when DNA
is present within the channel. Such effect is typically associated with the occlusion
due to DNA that diminishes the ionic fluxes.
Vice versa, current enhancement is also possible in a low-concentration electrolyte. 
Such effect can be ascribed to the excess of charge carriers accompanying DNA
as it occupies the pore region, that are loosely bound to the biomolecule and therefore 
available to conduction.
Both blockage and enhancement are observable in experiments and
their occurrence depends on the molarity, the degree of confinement, 
the nanopore material (such as $SiN$ or $SiO_2$), and possibly other 
physico-chemical parameters \cite{dekkergen,luan2013electro}.

In the present paper, we analyze how current modulation depends by several parameters
by employing a dual theoretical/computational approach. Due to the high level of confinement, 
one usually ignores the presence of convective currents, that is,
considers frictional forces strong enough to effectively damp out 
electro-osmotic effects. To test such an issue we shall compare results obtained by 
taking into account convective currents with those that exclude these contributions. Our description is encoded by the well-known 
Poisson-Nernst-Planck (PNP) theory \cite{zheng2011poisson}, as largely employed in the study of ionic transport in 
biological ion channels, such that only electrostatic, diffusive and migration
contributions are considered. In addition, in order to simplify the analysis
an effective one-dimensional equation based on a suitable
homogenization procedure is written down \cite{marconi2013ionic}. 
By identifying the entropic forces arising from the degree of confinement and, 
by imposing the local electroneutrality condition,
the conductance of the pore-DNA system can be estimated.
The enhancement/blockage diagram of the system reveals how the ionic current is modulated
as a function of molarity, pore geometry, effective charge on DNA and pore surface.
In addition, besides the role of the DNA intruder in the pore, the equations allow
determining the contribution of the reservoirs, with the related access resistance, 
to the overall transport.
Being a one-dimensional differential equation, the analysis is particularly manageable in terms of 
computational and modeling efforts. For instance, we will represent DNA as either 
a smooth or a corrugated cylinder, being either at rest or in motion within the pore,
and assess how the enhancement/blockage diagram varies correspondingly.
 With respect to our previous study \cite{marconi2013ionic}, we have improved the solution by replacing 
the local electroneutrality approximation used to determine the electric field
by the more refined solution of the Poisson equation.
The computational cost of the approach is modest and can serve as a useful 
method to perform an early survey of DNA/pore systems.

The paper is organized as follows in section \ref{Model} we define the model and its governing one
dimensional effective equations. In section \ref{Results} we perform numerical calculations and discuss the results concerning the conductance and a conductance diagram obtained by considering the variation of the conductance upon inserting 
into the pore a long thin cylindrical charged  object mimicking the DNA molecule. 
Finally, in section \ref{Conclusions} we present a few concluding remarks.

\section{Model system and reduction to a one-dimensional effective problem}\label{Conduttanza canale}
\label{Model}

\begin{figure}[htb]
  \centering
  \includegraphics[width=1.\textwidth]{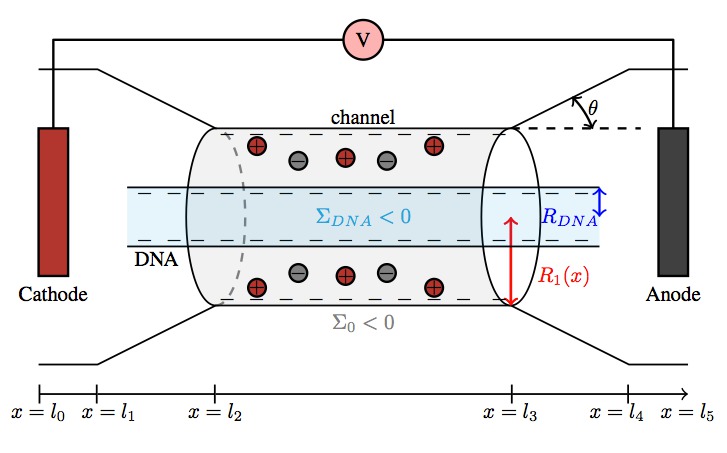}
  \caption{Geometry of the pore region.}
  \label{fig:Schema}
\end{figure}

\noindent  
We start by  describing the system and how it is modelled in effective one-dimensional terms.
The geometry of the three dimensional pore is sketched in Fig. \ref{fig:Schema} and consists of two  conical 
funnels connecting a straight pore.
The radius of the channel, which has cylindrical symmetry, can be written as a function of $x$ alone:

\begin{equation}
R_{ch}(x) = \begin{cases} 
R_0              & \mbox{if } l_0 \le x < l_1 \\ 
R_0 - B_0(x-l_1) & \mbox{if } l_1\le x \le l_2 \\
R_1              & \mbox{if } l_2<x<l_3 \\
R_0 + B_0(x-l_4) & \mbox{if } l_3 \le x \le l_4 \\
R_0              & \mbox{if } l_4<x \le l_5
\end{cases}
\end{equation}

\noindent The two conical funnels of variable radius  are joined by the channel with section of constant radius 
$R_1=R_0-B_0(l_2-l_1)$ and length $L=l_3-l_2$, whose inner walls carry a fixed surface 
charge of area density $\Sigma_0$, $B_0$ is the slope of the funnel 
and moreover $d=l_2-l_1=l_4-l_3$ and $\delta=l_1-l_0=l_5-l_4$.
Each funnel connects the pore to a reservoir
 and $l_{2,3}$ are the coordinates of the inlet and outlet of the cylinder, respectively.
  The DNA molecule is described as a very long thin
 cylinder of radius $R_{DNA}$ and with $R_{DNA}(x)<R_1<R_0$,
 whose axis coincides with the one of the pore \cite{marconi2013ionic}.  
 The  space between the pore walls and the DNA is filled by an electrolytic solution
  and   the cathode  and the anode are placed at $x=l_0$ and $x=l_5$,  respectively,  and a potential difference 
    is applied between them, as shown in Fig. \ref{fig:Schema}.

\noindent The evolution of the ionic concentrations is described
within the framework of  the Poisson-Nernst-Planck theory,
which combines  Fick's diffusion law with the  drift induced by the electric field and the convection of the ions
due to the motion of the solvent determined by the DNA movement  \cite{vlassiouk2008nanofluidic}.
The PNP three-dimensional model can be solved numerically, however  an effective one-dimensional
description can shed some light and has gained increasing popularity in recent times \cite{schuss2001derivation,
cervera2005poisson}.
In situations of interest the geometry of the channels is such that it is possible 
to neglect the variations of the
relevant fields along some directions and reduce the description  to a one-dimensional problem. 
In nanochannels of constant section, whose transverse width is much shorter than the longitudinal size, the reduction is straightforward,
 otherwise one must take into account the effect of the varying shape.

The lengths which determine the electrostatic properties of the system are the Debye screening length
, $\lambda_D$,  the Bjierrum length $\lambda_B$, the typical radius of the pore, of DNA
and the pore length.
The Debye length is a function of
the bulk densities of each ionic species, $n_b$, the dielectric permittivity, $\epsilon$, and the temperature $T$,
the electronic valence $z$ according to $\lambda_D= \sqrt{\frac{\epsilon k_B T  }{(ze)^2 }\frac{1}{ 2 n_b}}$,
where $k_B$ is the Boltzmann constant and $e$ the elctronic charge.
 The Bjerrum length is $\lambda_B=\frac{e^2}{4\pi\epsilon k_B T}$
 and under standard ambient conditions and for aqueous solutions is typically of the order of $0.7$ nm,
 while $\lambda_D$ in electrolytic solutions
of concentrations in the range $0.1-1.0$ M, $\lambda_D$ varies between $\simeq0.961$ and $0.304\,nm$.
Thus in channels of nanometric diameter $\lambda_D$ can exceed  their transverse size
and the ensuing behaviour of the solution is mainly determined by the presence
of the surface charges due to the partial overlap of the double layers of different bounding surfaces.
 One finds that the resulting ionic atmosphere is mainly
composed by counterions, a feature of capital importance because it yields the possibility of controlling through the surface
charge the ionic currents through the pores.

\noindent
In the present treatment we only consider   the steady current regime which does not require
 the full time dependent solution of the PNP equations. The condition that
the ionic current densities, $J^\pm$, must have vanishing divergence, $\nabla \cdot {\bf J}^\pm ({\bf r}) =0$,
together with the property of impenetrability of the walls,
yield the following relation between the average axial 
component, $J_x$, of the three-dimensional  current and the variable section of the pore:
\begin{equation}
\label{eq:flux axial direction}
I^{\pm}=\langle J_x^{\pm}(x) \rangle S(x)=constant
\end{equation}
where $S(x)$ is the transverse section of the system and $I^\pm$ the current.
After this premise,
in the stationary state $I^\alpha$ is  the sum of four different terms:
the diffusive contribution, the entropic term, accounting for the
modulation of the confinement  \cite{jacobs1967diffusion,zwanzig1992diffusion,reguera2001kinetic}, the migration contribution stemming from the driving electric field
and a convective term due to the electro-osmotic flow induced by the  charges present on the pore and DNA surfaces. The r.h.s.
of the following expression encodes the four different contributions,
\trr{
\begin{eqnarray}
\label{eq:eqprimoordine}
\frac{d c^{\pm}(x)}{dx}-\frac{d\,ln\,S(x)}{dx}c^{\pm}(x)-\frac{e z^{\pm}\langle E(x) \rangle}{k_B T}c^{\pm}(x)\nonumber\\
+c^{\pm}(x)  \frac{v_{conv}}{D^\pm}= -\frac{I^{\pm}}{{D^\pm}}
\end{eqnarray}
}
where $D^{\pm}$ is the species coefficient of diffusion.

\trr{The convective velocity $v_{conv}$, will be discussed below,  depends
on the surface charges of the pore and DNA  and on the electric field along the axis direction. }
Equation \eqref{eq:eqprimoordine} represents an ordinary first order differential equation for the one-dimensional linear density 
$c^{\pm}$ of species $\pm$, which is  related to the three dimensional ionic densities $n^\pm (\rr,t)$
by  the following sectional averaging:
\begin{equation}
c^{\pm}(x,t)=\langle n^{\pm}(x,t) \rangle S(x)= \int_{S(x)} dS n^\pm (\rr,t)
\label{averagec}
\end{equation}
where the sectional area $S(x)$ for the system shown in Fig. \ref{fig:Schema} is
\begin{equation}
S(x)\equiv\pi\Bigl(R_{ch}^2(x)-R_{DNA}^2(x)\Bigr)\, .
\end{equation}

\noindent
Similarly,
we define the sectionally averaged electric field:
\begin{equation}
\langle E(x,t) \rangle \equiv \frac{1}{S(x)}\int_{S(x)} dS E_x (\rr,t)\, ,
\end{equation}
which as shown in ref. \cite{marconi2013ionic,gillespie2002coupling} satisfies the following differential equation
\begin{eqnarray}
\frac{d \langle E(x) \rangle }{dx}+\frac{d\,ln\,S(x)}{dx}\langle E(x) \rangle\nonumber\\
=\frac{1}{\epsilon S(x)}\Bigl[z^+ c^+(x)+z^- c^-(x)+2\pi R_{ch}(x)\Sigma_w(x)\nonumber\\
\times\sqrt{1+\Bigl(\frac{dR_{ch}(x)}{dx}\Bigr)^2}+2\pi R_{DNA} \Sigma_{DNA}\Bigr]
\label{eq:campoelettrico}
\end{eqnarray}
where $\Sigma_w(x)$ is the wall charge density, being $\Sigma_0$ in the pore and
zero elsewhere.

The terms proportional to the surface charge of the walls $\Sigma_0$ and of the DNA intruder
$\Sigma_{DNA}$ appear as source terms in the one-dimensional representation 
and stem from the boundary conditions and  on  the reduction 
of the original three dimensional problem.
The  geometrical term containing the logarithm of $S(x)$
 enforces the conservation of the flux of the
electric field in a system of variable  pore section.

The set of effective one-dimensional equations here considered represents a convenient approach to the solution of full three-dimensional problems, which involve a demanding numerical effort.
As far as the performance of the one-dimensional PNP model (PNP1) versus the corresponding three dimensional  model (PNP3)
we verified that the narrower the channel, the larger is the difference in profiles between the two types of solutions \cite{marconi2013effective}.

In the case of a binary electrolytic solution  one has to solve three coupled differential equations:
the two  equations \eqref{eq:eqprimoordine} are needed to determine the local concentrations of the $\pm$ species,
while the third equation \eqref{eq:campoelettrico} determines the electric field produced by the
ionic charge distribution and the fixed charges.  In the literature one often employs the so-called local electroneutrality approximation (LEN)  \cite{kontturi2008ionic,dickinson2011electroneutrality} because it avoids the effort of finding the solution of  eq.\eqref{eq:campoelettrico} which is not known in closed analytical form.
The LEN approximation states that locally there is no charge separation, a fact that may be justified because the typical values of the screening length
$\lambda_D$ are small as compared with the  system dimensions, such as the longitudinal and transverse channel sizes.
 In other words one assumes that at each point $x$ the mobile
charges exactly balance the fixed surface charges, as it occurs in the limit of vanishing $\lambda_D$.
The electric field does not satisfy the one dimensional Poisson equation \eqref{eq:campoelettrico},
but takes values that enforce the absence of charge separation.
 Under steady state conditions one can express the charge
density distribution in terms of the fixed surface charge
distribution and obtain a very simple equation for the density profiles in terms of the electric current and the mass
current  \cite{marconi2013ionic}.
On the other hand,
 dimensional analysis shows that the LEN may break down in nanometric systems, that is, when 
 the system size is of the same order as $\lambda_D$.

To fix 
the convective velocity, $v_{conv}$, that
plays the role of an input parameter in the PNP model,  we follow Ghosal's  treatment
\cite{ghosal2007effect,ghosal2012electrokinetic,ghosal2006electrophoresis}.
The presence of surface charges and of the applied electric field, $E$, along the axis,
 generates an electroosmotic flow (EOF), whose velocity is parallel to $E$. The
 resulting radial velocity profile, $u(r)$ of the fluid, vanishing at the pore surface, reads:
\begin{equation} 
\frac{u(r)}{u_{o}} =  \frac{\phi(r) - \phi_w}{\phi_w} 
\label{velocity2}
\end{equation}
where the potential $\phi(r)$ is the solution of the two dimensional Debye-Huckel equation
in the anular region $R_{DNA}\leq r \leq R_1$: 
\begin{equation} 
\phi(r) = \frac{\Sigma_{DNA} \lambda_D}{ \epsilon}  \left[ A I_{0} \left( \frac{r}{\lambda_D} \right)  + B K_{0} \left( \frac{r}{\lambda_D} \right) \right] ,
\label{bessel2}
\end{equation} 
$\phi_w$ is the value at $r=R_1$,
 $u_{0} = \epsilon E \phi_w / \mu$ a characteristic electroosmotic
velocity and $\mu$ the dynamic viscosity of the solution. $K_{0}$ and $I_{0}$ are modified Bessel functions of integral order. 
The constants $A$ and $B$ are determined (see ref. \cite{ghosal2007effect}) by imposing the boundary conditions
\begin{eqnarray} 
-  \epsilon  \phi^{\prime}(R_{DNA}) &=& \Sigma_{DNA}
\label{bc1}\\
 \epsilon  \phi^{\prime}(R_1) &=& \Sigma_0 .
 \label{bc2}
\end{eqnarray} 
We approximate the convective
 velocity featuring in eq. \eqref{eq:eqprimoordine}  by the following sectional average:
 \begin{equation}
v_{conv}  = \frac{2}{R_1^2-R_{DNA}^2} \int_{R_{DNA}}^{R_1} dr r u(r) .
\label{media}
\end{equation} 
Notice that in the cases here considered such a velocity turns out to be larger than the translocation velocity of the DNA molecule
 whose value can be estimated as:
$$
v_{DNA}= u_o  \frac{\phi(R_{DNA}) - \phi_w}{\phi_w} .
$$

\trr{In order to give an idea of the importance of the electric current associated with the EOF, $I_{EOF}$,
with respect to the conduction current, $I_{Ohm}$, we  consider the ratio :
$$
 \frac{I_{EOF}}{I_{Ohm}}
\approx\frac{k_B T}{e^2 \mu D} \Sigma_0^2\lambda_D^2 \lambda_B .
 $$
 For sodium ions in water at  1 M concentration the above ratio is approximately $0.003$, where $D$ and $\mu$ are the diffusion coefficient of the ions and the dynamic viscosity of water, respectively.  }
 
As mentioned in the introduction
to detect the passage and eventually sequence DNA, one monitors the ionic conductance $G$ of the pore.
On theoretical grounds,
the total resistance $\Re_{tot}$ of the system shown in Fig. 1 is the sum of the  resistances of  five
different pieces: the resistance associated with the two funnels plus the resistance due to the two cylindrical regions adjacent the electrodes and 
the central part, 
 $\Re_{tot}=\Re_{ch}+2\Re_{cone}+2\Re_{reservoir}$. 
In the absence of surface charges 
the total conductance can be approximated by the following  Ohmic 
formula (see ref. \cite{marconi2013ionic}) :
\begin{equation}
G=\frac{\pi}{\rho_0}\frac{1}{\frac{L}{R_1^2}+ \frac{2 \delta}{R_0^2}+ \frac{2}{B_0}(\frac{1}{R_1}-
\frac{1}{R_0})}
\label{condcanalelibero}
\end{equation}
where
 $\rho_0$ is the resistivity defined in terms of the coefficient $D$ as $\rho_0=\frac{k_B T}{D}\frac{1}{2n_b e^2}$.
We first evaluate the effect of  the presence of a cylindrical intruder of radius $R_{DNA}$ 
within the pore in the limit of  $R_0>>R_1$ and $L>> R_1$ and no charges.
By taking into account the reduced section available  one estimates the difference in conductance
between the free and the  obstructed pore we find from eq. \eqref{condcanalelibero}:
\begin{equation}
G(R_{1})-G\bigl(\sqrt{R_{1}^2-R_{DNA}^2}\Bigr)\simeq\frac{\pi}{\rho_0}\frac{1}{L}R_{DNA}^2  \,.
\end{equation}
Although
the  conductance due to the obstruction  always results negative on the basis of steric arguments,
the presence of  negative  charges on the DNA and not included in 
formula  \eqref{condcanalelibero}, may lead to a different answer. In fact,
some positive ions migrate from the reservoirs towards the negative surfaces and increase locally the number of mobile
carriers and
one observes a conductivity larger than in the absence of DNA.
Such a phenomenon is particularly relevant at low ionic concentrations and is related
to the so-called surface conduction mechanism \cite{bocquet2010nanofluidics,lyklema1998surface}.

The competition between enhanced conduction due to the presence of surface charges 
and the depletion due to the reduction of the available pore section due to the DNA
leads to an interesting dependence of the relative conductance 
on the DNA features. 

Hereafter, we shall study the conductance variations, as the concentration of the
solution and the surface charge are varied,
by comparing the numerical results relative to the one-dimensional PNP model with the predictions based  on the LEN  theory, which provides simple expressions
for small potential and concentration drops.


\section{Methods and numerical results }
\label{Results}

The simulations were performed considering
systems comprised of the five different regions illustrated in  Fig. \ref{fig:Schema}.
In the numerical calculations $R_1$ took values  between $5\,nm$ and $20\,nm$, $B_0 = (R_0 - R_1)/d$ 
while  the other geometrical parameters were always $R_0= 25 nm$  $L=34\,nm$, $\delta=3\, nm$ , $d=20\,nm $.
The DNA is modelled as a long rigid cylinder of radius $R_{DNA}=1.1\,nm$, coaxial to the  nanochannel, and having surface charge density  $\Sigma_{DNA}$. 
This schematic representation of DNA is justified by the persistence length of the dsDNA molecule which is about $50\,nm$.
 The surface charge densities were assumed
to have values $\Sigma_{0}=-0.375 \, e/nm^{2}$, $\Sigma_{DNA}=-0.38175\, e/
nm^{2}$, $R_{DNA}=1.1\, nm$.
 A voltage drop was applied on the ends of the system:
\begin{equation}\label{eq:phi agli estremi}
 \phi(l_0)=\phi_I\,\,\,\,\,\,\,\,\,\phi(l_5)=\phi_O .
\end{equation}
In order to solve numerically eqs.  \eqref{eq:eqprimoordine}  and   \eqref{eq:campoelettrico}   we have introduced a one-dimensional
mesh $\Delta x$ 
 and defined non dimensional quantities in the following way:
the Debye length, $\kappa^{-1}=\tilde{\kappa}^{-1}\Delta x$,
the concentration, $n=\tilde{n}\Delta x^3$,
the charge density $\Sigma=\tilde{\Sigma}\Delta x^2/e$,
the electric potential $\psi=e \tilde{\psi}/k_B T$,
the applied potential difference $\Delta V=e\tilde{\Delta V}/k_B T$.

\trr{In the following, unless explicitly stated we shall perform calculations by neglecting 
the convective contribution to the current
and include this term only when computing the global phase diagram.}

We started by measuring the dependence of the total electric current, $ I_{tot}=I^- - I^+  $, on the applied potential difference
$\Delta V$ for various values of the salt concentration
relative to  a solution of ions of identical masses and mobilities and extracted the
conductance $G$.
In the case of the empty pore we found   the $I-V$ characteristic to be linear  at 
 concentrations $0.01\,M$,
  $0.1\,M$,$ 1.0\,M$  (data not reported).
   On the other hand, the  conductance  as concentration is lowered
  at first decreases almost linearly,
 showing a standard bulk Drude behavior, but finally reaches a plateau value  when $n_b$
 is of order of the  ratio $\Sigma_0/R_1$.

\begin{figure}[htb]
  \begin{center}
  \includegraphics[width=1.\textwidth]{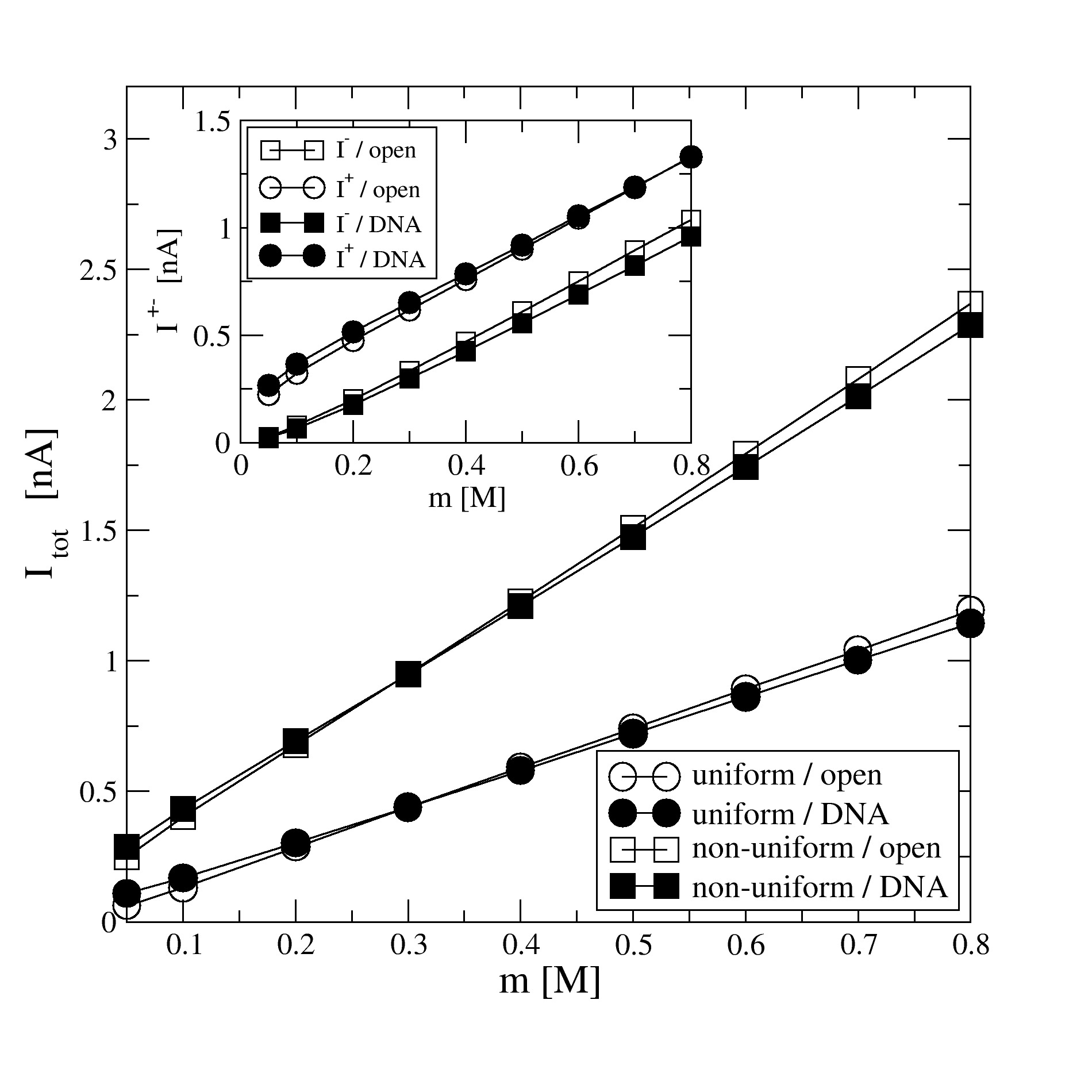}  
  \caption{Comparison between the total electric currents in the case of  the double funnel 
  geometry (squares) and a straight cylindrical geometry (circles). 
  Inset: Individual ionic currents for  the double funnel geometry. Circles refer to the case where DNA is present,
  while squares to the free case.
   Data refer to a channel radius $R_1=5\,nm$, length $L=34\,nm$ and potential difference $0.2\,V$. 
  The wall and DNA surface charge densities are $\Sigma_{0}=-0.375\,e/nm^2$ and $\Sigma_{DNA}= - 0.38175\,e/nm^2$, 
respectively and \trr{no convection $v_{conv}=0$.}}
  \label{fig:corr tot omogeneo VS disomogeneo}
  \end{center}
\end{figure}

In order to understand the effect of the two vestibules we compared 
 the results relative to the set up of Fig. \ref{fig:Schema} with those relative to a system with 
 constant radius $R_0=R_1$,
that is a straight cylindrical pore.
Fig. \ref{fig:corr tot omogeneo VS disomogeneo} displays the ionic current  versus molarity in these two cases
for the empty system and for the system partially occluded by DNA   
when $R_1=5\,nm$.
We observe that the total ionic current in the non uniform channel,  by virtue of its larger cross section, is larger than the one relative to  the straight model.
In the inset of Fig. \ref{fig:corr tot omogeneo VS disomogeneo} the currents of counterions and coions are reported separately,
limited to the case of the
inhomogeneous pore. One can observe  that the counterion current, $I^+$, is always larger 
 in the presence of DNA than in the free-DNA case due to the surface induced enrichment
 and that the difference increases at low molarities where the surface conduction mechanism is more evident.
On the contrary, the coion current, $I^-$,  is sensibly higher for the free-DNA case at large molarities
where the steric hindrance overwhelms the Coulombic repulsion, but the difference disappears as the Debye length
increases.

Variations in conductance associated with the presence of DNA are best appreciated
by considering the relative conductance deviation defined as the ratio:
\begin{equation}\label{conduttanza relativa}
  \Delta G=\frac{G_{DNA}-G_{free}}{G_{free}}
\end{equation}
where $G_{free}$ is the conductance of the free pore, while $G_{DNA}$ is the same quantity in the presence of DNA.

\begin{figure}[htb]
  \begin{center}
  \includegraphics[width=1.\textwidth]{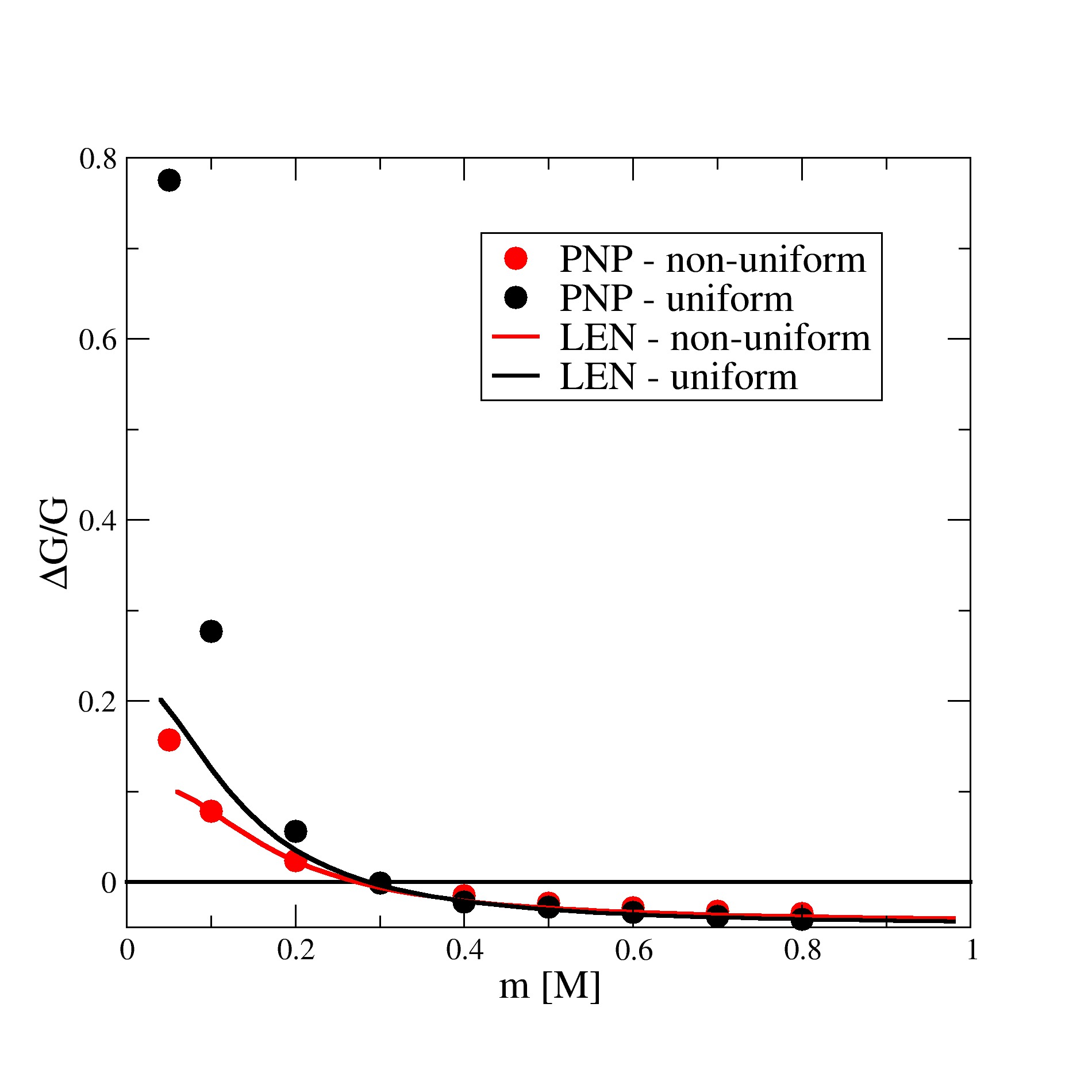}
  \caption{Relative conductance in the presence of DNA intruder. Comparison between  the relative conductance versus molar concentration of a cylindrical channel  and the 5 stages system of fig. \ref{fig:Schema}, corresponding to
  a radius of the channel $R_1=5\,nm$ and \trr{without convection ($v_{conv}=0$).
   The wall and DNA surface charge densities are the same as in Fig. \ref{fig:corr tot omogeneo VS disomogeneo}  }.  The continuous lines represent the corresponding results obtained within the LEN approximation.
 In the case of the uniform geometry
  the LEN gives a lower value of $\Delta G$, whereas for the system B the same quantity is larger.}
  \label{fig:conduttanzaconfronto1}
  \end{center}
\end{figure}

\begin{figure}[htb]
  \centering
\includegraphics[width=1.\textwidth]{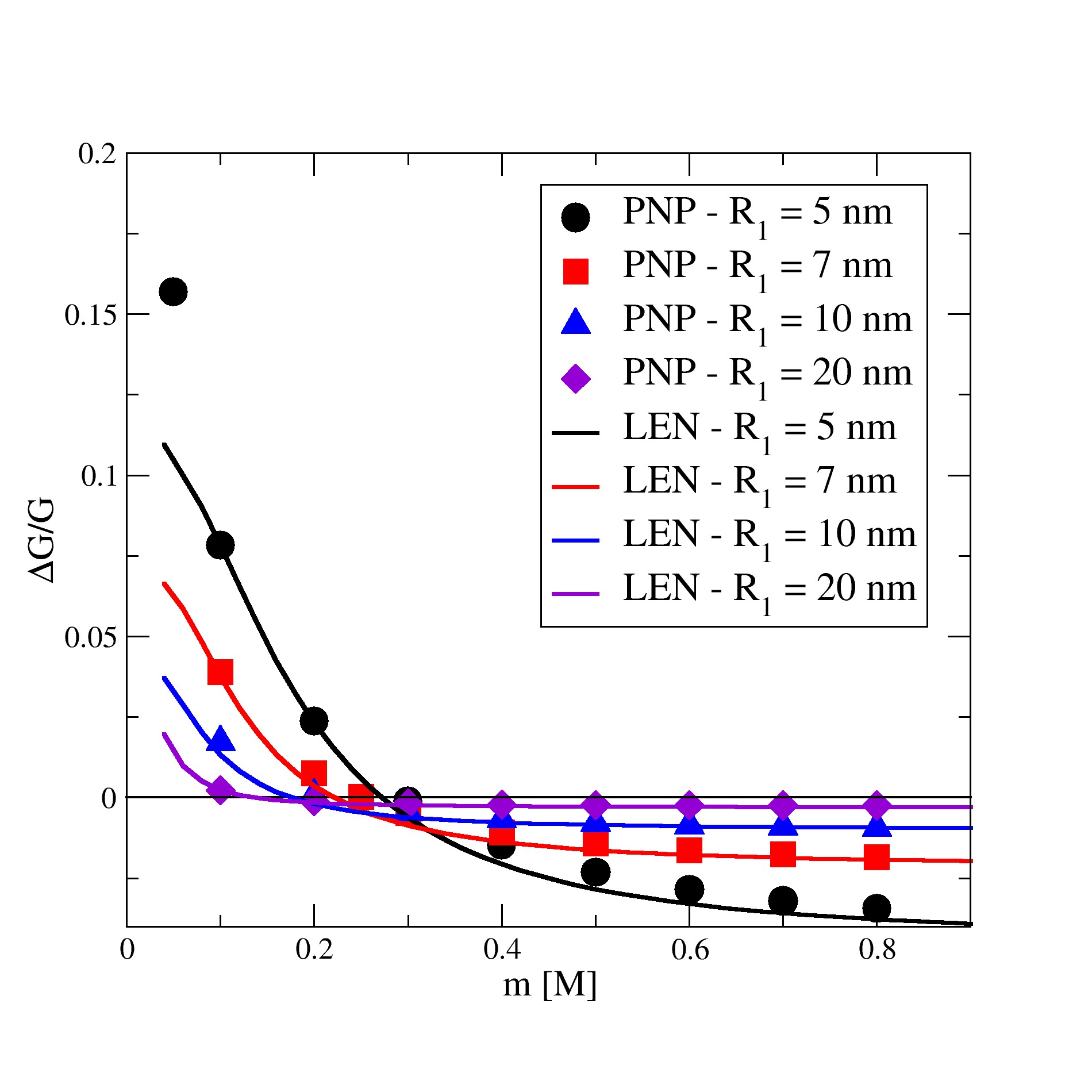} 
  \caption{Relative conductances versus bulk ion concentration for channel radii $R_1$ between $5\,nm$ e $20\,nm$
  computed within the PNP one dimensional equation and  \trr{without convection ($v_{conv}=0$)}.   The surface charge density of channel wall is $-0.375   \,e/nm^2$ \trr{and $\Sigma_{DNA}=-0.38175\,e/nm^2$}. The continuous lines report the corresponding LEN approximation results.}
  \label{fig:conduttanzaconfronto2}
\end{figure}

\begin{figure}[htb]
\centering
\includegraphics[width=1.\textwidth]{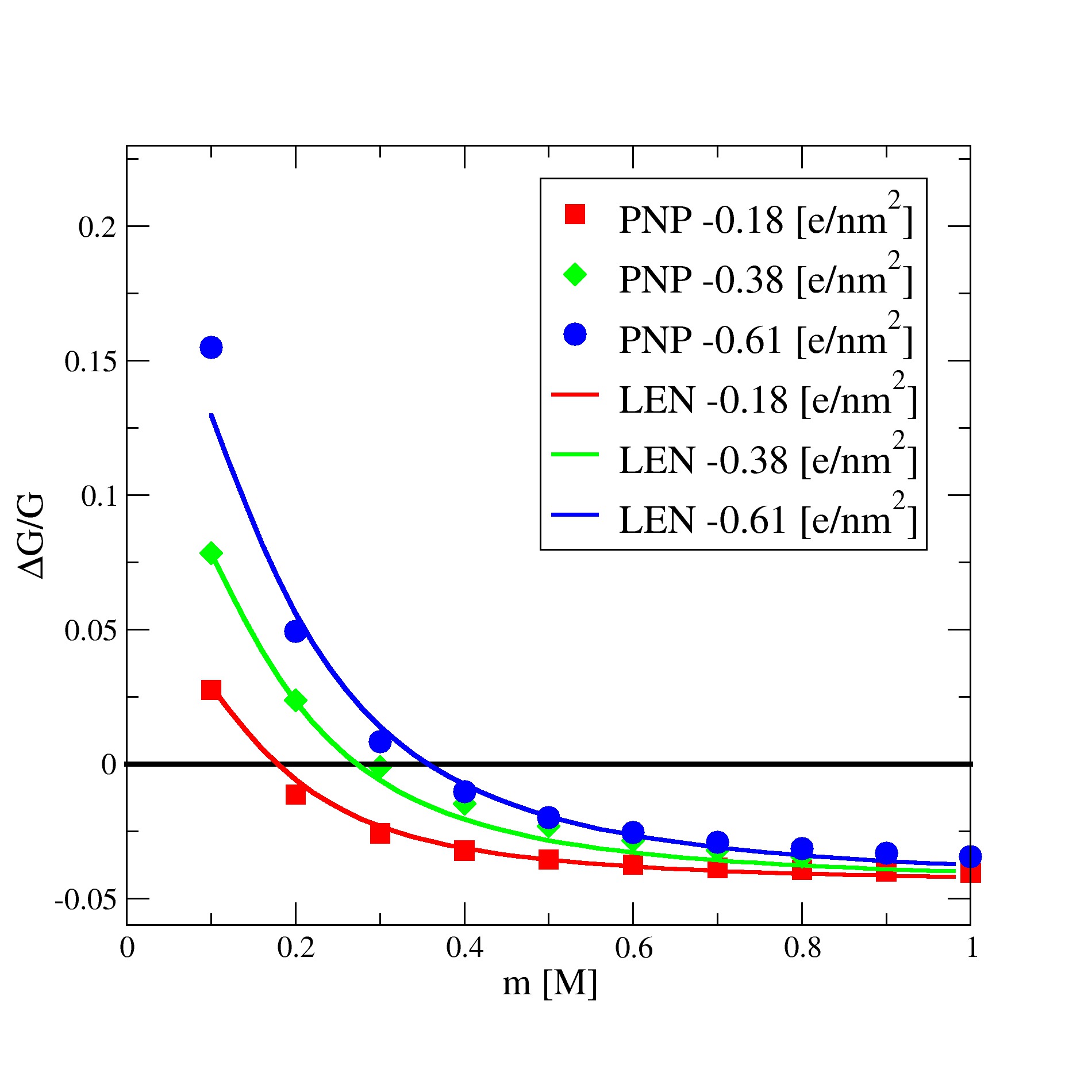} 
\caption{ Relative conductance against bulk ion concentration. The channel radius is $5\,nm$ ,
\trr{$\Sigma_{0}=-0.375\,e/nm^2$ } and the surface charge concentration density of DNA varies between $-0.61\,e/nm^2$ and $-0.18\,e/nm^2$
\trr{and $v_{conv}=0$}. 
 The continuous lines report the results obtained via the LEN approximation.}
\label{fig:conduttanzacaricaA}
\end{figure}

\begin{figure}[htb]
\begin{center}
\includegraphics[width=1.\textwidth]{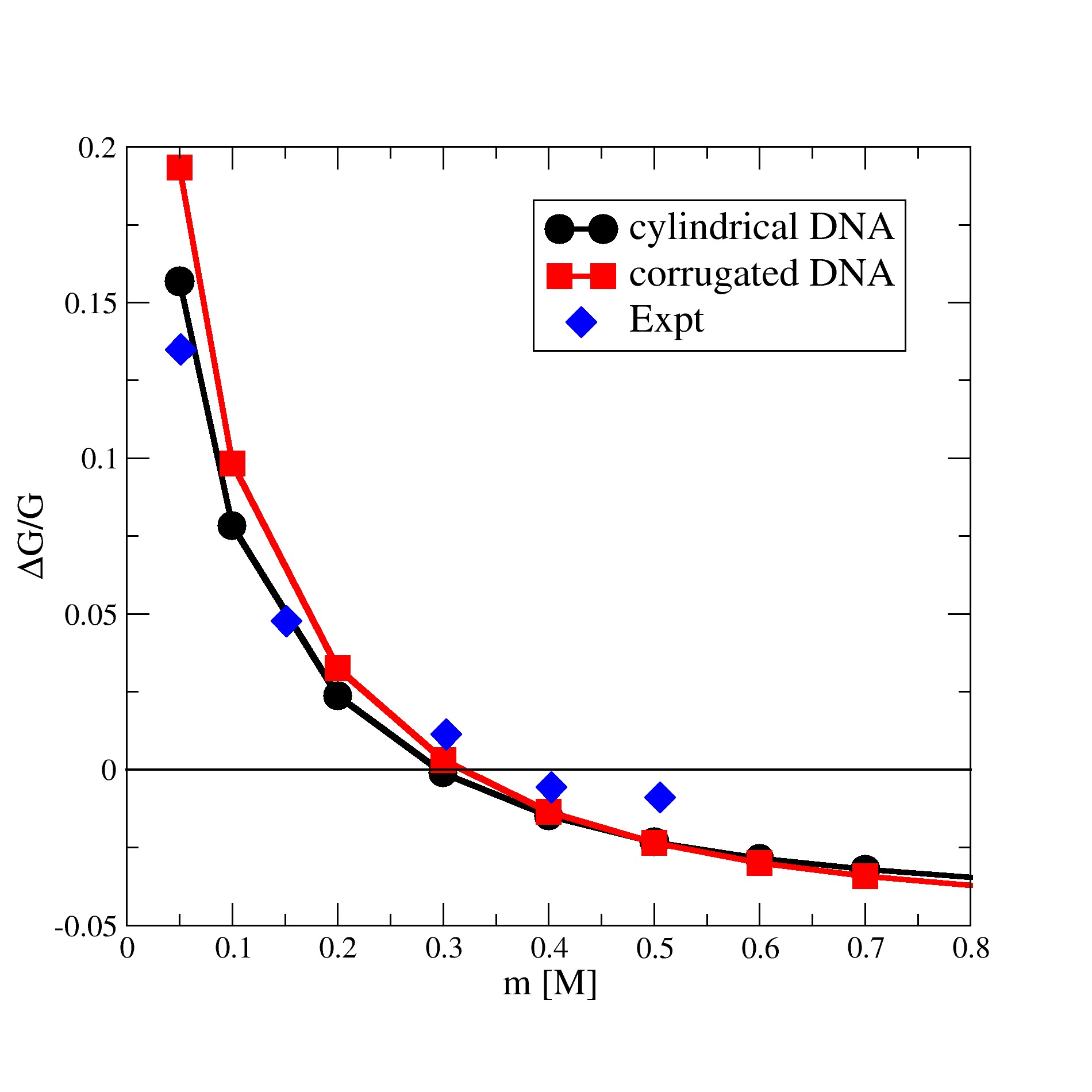}
\caption{Relative conductance versus  bulk ion concentration for a pore
of radius $R_1=5\,nm$.  \trr{The values of the surface charges are the same as in Fig \ref{fig:corr tot omogeneo VS disomogeneo}.}
Circles indicate the PNP results relative to the cylindrical DNA, squares those
corresponding to the model with corrugations, while diamonds are the experimental data of Smeets et al .\cite{smeets2006salt}. 
}
\label{corrugato}
\end{center}
\end{figure}

\noindent
Fig. \ref{fig:conduttanzaconfronto1} illustrates the behavior of such an observable 
as the electrolyte concentration
is varied. When the molarity is below a certain threshold value, $M_c$,
one observes a positive value of $\Delta G/G$, which corresponds to an enhanced  conductance 
with respect to the free-DNA case.
Above $M_c$, instead, the presence of the DNA intruder partially hinders the passage of ions
and determines a reduction of the conductance with respect to its open pore value. 
Following the name convention the two regimes are named
enhancement and blockage, respectively.
We have also tested
the importance of the geometry  by comparing 
the relative conductance of a straight cylindrical pore with that relative to the double funnel-cylinder system.
The effect of the inhomogeneity appears  to be relevant only at low concentrations
($<0.2 M$) while the critical concentration, $M_c$, results nearly independent on the channel geometry.
Fig. \ref{fig:conduttanzaconfronto1} seems to indicate that  the straight pore geometry
is more sensitive to the presence of DNA since
the relative conductance varies faster with decreasing molarity.
We ascribe this feature to the fact that the absolute value
of its conductance is lower than the one associated with the double funnel, while the
variations due to the DNA intrusion are comparable in the two situations.
In Fig. \ref{fig:conduttanzaconfronto1} we also reported  the results obtained using  the LEN approximation,
where one can appreciate
that at large molarities the  agreement between the data obtained in the present work and the LEN
is fairly good, whereas at low molarities
corresponding to larger Debye lengths  the LEN theory tends to underestimate the difference in conductance between the DNA-free and DNA case.

The importance of the radius of the pore is stressed in
Fig.~ \ref{fig:conduttanzaconfronto2}, where
 the relative conductance is shown for  $R_1=10,20$,
 for a reduction of   $38 \%$ of the nominal charge DNA density,
  while the remaining parameters were the same as in Fig.
  \ref{fig:conduttanzaconfronto1}. 
 One can see not only that the larger the radius the smaller the crossover concentration,
 but also that the sensitivity decreases with increasing size as one can see from the fact that
 for a fixed value of the molarity  the relative conductance of the
 corresponding to $R_1=20\, nm$ is in general lower than the conductance relative to $R_1=10 \,nm$.
  For the sake of comparison, we also display the  corresponding results  of the  LEN approximation
  and remark that for  the crossover value appears to be underestimated by the LEN, which predicts 
that the blockage  extends to lower molarities with respect to the PNP.
\begin{figure}[htb]
\begin{center}
  \includegraphics[width=1.0\textwidth]{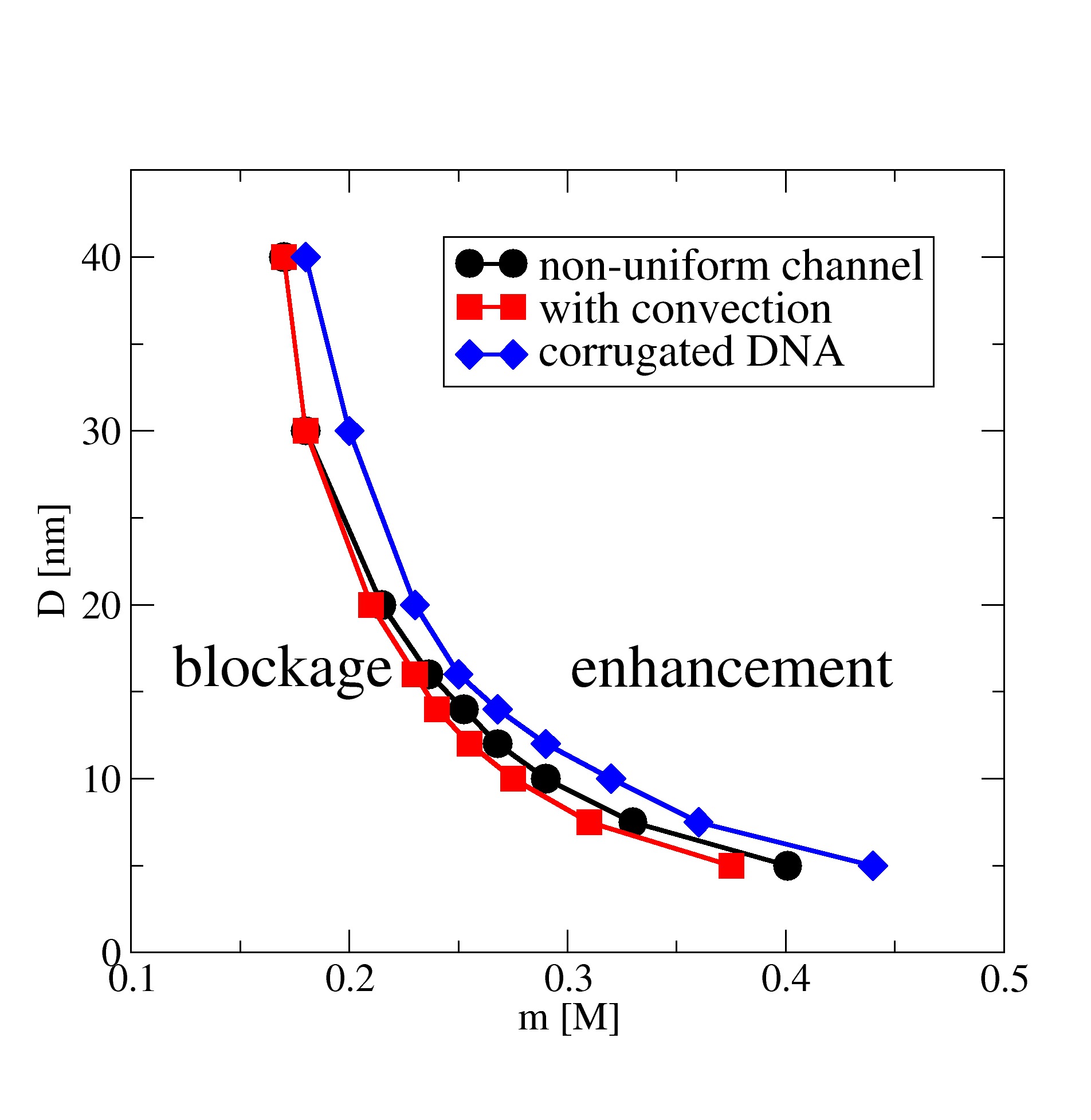}
        \caption{
  Dependence of the critical concentration on the nanochannel diameter and bulk ion concentration. The surface charge density of nanochannel is $-0.375\,e/nm^2$ and $\Sigma_{DNA}=-0.38175\, e/
nm^{2}$  . Three cases are presented: non uniform channel (circle), DNA convection (square), corrugated DNA (diamond). The convective term $v_{conv}$ has been obtained by using eq. \eqref{media}.
\trr{ In the corrugated case $R_{DNA} (x)$ was given by eq.\eqref{DNAcorrugato}
with $r_0 = 1.1\,nm$ , $A=0.5\,nm$ and $P=3.4\,nm$. }
}
  \label{fig:diagfase}
\end{center}
\end{figure}

 Fig.~\ref{fig:conduttanzacaricaA} illustrates numerical results for $\Delta G$ for a non uniform channel of radius $R_1=5\, nm$
versus salt concentration,  varying in the interval
$0.05\,M$ a $1\,M$, corresponding to Debye lengths ranging from $0.3\,nm$ to $1\,nm$,
 for different choices of the value of the DNA charge (that is  for a reduction of  $0 \%$, $38 \%$ e $70 \%$ of the nominal charge DNA $-0.61\,e/nm^2$) : as the DNA charge
 decreases the crossover  point moves towards lower concentrations in agreement
 with the fact that the conductance is dominated by the surface conduction mechanism according to which
 a larger  $\Sigma_{DNA}$ determines an enrichment of the counterions and thus a larger conductance.
 In this case the LEN approximation also appears to work quite well as compared to the PNP method.

An interesting aspect regards the role played by the roughness of the surface of the DNA molecule. 
So far, we represented the DNA as a uniform cylinder of radius $R_{DNA}$ and neglected 
its double helix structure. We, now, consider DNA as having a corrugated shape of cylindrical symmetry 
and radius varying along the symmetry axis according to the law:
\begin{equation}\label{DNAcorrugato}
  R_{DNA} (x) = r_0 + A\sin (\frac{2\pi}{P}\, x)
\end{equation}
where $r_0 = 1.1\,nm$ , $A=0.5\,nm$ and $P=3.4\,nm$. 
Consistently, eqs. \eqref{eq:eqprimoordine} and \eqref{eq:campoelettrico} are modified in order to take 
into account the variation of $R_{DNA}$ with position.
Fig. \ref{corrugato} shows that the effect of the corrugation is appreciable only in the low
concentration regime where the surface conduction dominates and is influenced by $\Sigma_{DNA}$ and the currents are more sensitive 
to the geometrical details of the charge distribution.  The same figure also shows the 
comparison between the theoretical results and the 
corresponding experimental result of Dekker and coworkers \cite{smeets2006salt}. 


A global picture of the conducting properties of DNA-pore system is provided by the  two dimensional 
conductance diagram, whose axes are the molar concentration and the channel diameter as shown in
Fig. \ref{fig:diagfase}.
For each pore diameter, one determines the critical concentration $M_c$ where the relative conductance
 changes sign, as for instance shown in Fig. \ref{fig:conduttanzacaricaA}.
The plane shown  in Fig. \ref{fig:diagfase} is divided in two regions:
 above the line the ionic electric current is reduced since in wider pores the prevailing mechanism 
of conduction is Ohmic and the geometric effect  beats the surface conduction; below the line, instead, 
the extra charges made available to conduction within the pore by the presence of  fixed charges 
gives rise to the surface conduction and thus to an enhanced current.
Each point displayed in  Fig. \ref{fig:diagfase} represents the value of the critical concentration $M_c$
where the crossover  enhancement/blockage  occurs.
\trr{In Fig  \ref{fig:diagfase} for the sake of comparison we have included the effect  on the variation of the relative
conductance
of the convective term
using the value provided by formula \eqref{media}.} 
The three different lines refer to:
a) the channel shown in Fig. \ref{fig:Schema},
b) the same set up but with a corrugated DNA intruder as described by \eqref{DNAcorrugato}  and 
c) \trr{the effect of the convection velocity,
 $v_{conv} $, given by formula \eqref{media} on the phase diagram.}
\trr{Fig. \ref{fig:diagfase} also shows the effect of DNA convection, which is negligible 
for radii larger than $7.5\,nm$.  
Indeed,  one can observe that the inclusion of convection determines only a small shift of the "coexistence line".
}
On the other hand, Fig. \ref{fig:diagfase} also indicates that the conductance diagram is shifted 
in the direction of higher molar concentrations 
when one includes the possibility of DNA corrugation.
In addition, Fig. \ref{fig:diagfase}  shows that by increasing the pore radius 
the critical concentration decreases slightly and tends to be independent on the pore size.
 At high concentration the enhancement current can not be recognized. However, 
it is possible to detect the passage of DNA in the pore by the current blockage.

\section{Conclusions}
\label{Conclusions}
We have employed a simple one-dimensional representation of a pore-electrolyte-DNA system 
widely used in experiments on DNA translocation. Our focus has been 
in understanding how the ionic currents are altered by the presence of DNA in the pore,
modelled as a charged cylindrical intruder.
The analysis is based on a one-dimensional reduction of the three dimensional PNP model 
which treats the ions in the continuum and the electrostatic interactions in a mean field fashion.
The resulting effective equations are of diffusive type and display 
the presence of the so-called entropic term, stemming from the 
variations of the geometry of the channel along the direction of the axis, 
and of a driving electric field due to the charges.

We have found instructive to
 compare the results of one dimensional PNP model 
 with the corresponding
results of the LEN theory. The latter in spite of its simplicity can give a first
hint of the behavior of a complex system such as a pore.
The LEN calculation shows that the crossover  from blockage to enhancement is qualitatively reproduced,
but quantitatively its predictions  become more inaccurate especially at low molarities. 
It is evident that the LEN yields crossover values $M_c$
for pores of large radius which are too low.
We attribute this feature, to the fact that the LEN underestimates the effect of the
surface charges and thus determines a  lower conductance. 
In the future work the results of  this one dimensional model will be tested against full three dimensional 
Lattice Boltzmann simulations  \cite{melchionna2011electro,marini2012charge}. 

\newpage


\end{document}